\documentclass[aps,prl,twocolumn,superscriptaddress]{revtex4-2}
\usepackage[T1]{fontenc}
\usepackage{graphicx}
\usepackage{amsmath}
\usepackage{hyperref}

\newcommand{\pumpup}{\Gamma_{\uparrow}}
\newcommand{\pumpdn}{\Gamma_{\downarrow}}

\begin{document}

\title{Photon condensation from thermal sources and the limits of heat engines}

\author{Lu\'{\i}sa Toledo Tude}
\affiliation{School of Physics, Trinity College Dublin, Dublin 2, Ireland}
\affiliation{Institute for Cross-Disciplinary Physics and Complex Systems (IFISC) UIB-CSIC, Campus Universitat Illes Balears, 07122, Palma de Mallorca, Spain.}
\author{Emily Haughton}
\affiliation{School of Physics, Trinity College Dublin, Dublin 2, Ireland}
\author{Paul R. Eastham}
\affiliation{School of Physics, Trinity College Dublin, Dublin 2, Ireland}

\date{\today}
\begin{abstract}
The trapping and cooling of photon gases in microcavities has been used to create Bose-Einstein condensates. We investigate the conditions required for condensation in dye-filled microcavities, with photon populations created either by driving a transition of the dye, or by coupling the cavity modes to a thermal photon reservoir such as sunlight. We find that the threshold pump temperature, above which condensation appears, is determined by the second law of thermodynamics. The minimum achievable threshold is that of a reversible three-level heat engine, which we show arises in the dye-pumped case, and for pumping of the modes of a two-level cavity. For a many-level cavity condensation occurs at a similar but higher temperature. Our results show that photon condensates can be produced by pumping with incoherent thermal sources, opening possibilities for coherent light generation, energy harvesting, and experimental studies of quantum heat engines. 
\end{abstract}


\maketitle

The thermalization of cavity photons by inelastic scattering can be used to produce equilibrated photon gases, at densities and temperatures sufficient to achieve Bose-Einstein condensates. Such condensates have been observed in dye-filled microcavities~\cite{klaers_boseeinstein_2010,marelic_experimental_2015,greveling_density_2018}, plasmonic lattices~\cite{hakala_boseeinstein_2018}, doped fibers,~\cite{weill_boseeinstein_2019,weill_bose-einstein_2021} and semiconductor microcavities~\cite{schofield_boseeinstein_2024,pieczarka_boseeinstein_2024,barland_photon_2021}. In these systems photons are rapidly absorbed and re-emitted by the medium, allowing the gas to cool without significant loss of photons. The resulting condensates have been used to explore the physics of near-ideal Bose gases, including grand-canonical fluctuations~\cite{schmitt_spontaneous_2016,klaers_statistical_2012}, fluctuation-dissipation theorems~\cite{ozturk_fluctuation-dissipation_2023}, heat capacity~\cite{damm_calorimetry_2016}, equations-of-state~\cite{busley_compressibility_2022}, and the approach to the continuum limit~\cite{erglis_photonic_2024}. Non-equilibrium effects have also been studied, including the breakdown of thermalization~\cite{kirton_thermalization_2015,kirton_nonequilibrium_2013} and coherence~\cite{tang_breakdown_2024} and non-equilibrium phase transitions~\cite{ozturk_observation_2021,erglis_photonic_2024}. 

An important feature of these systems is that the photon distribution functions are close to thermal, due to the fast inelastic scattering. This distinguishes photon condensates from typical lasers~\cite{schmitt_thermalization_2015} or exciton-polariton condensates~\cite{gomez-dominguez_materials_2025}. Nonetheless, there is a loss of condensate photons through the cavity mirrors, which is offset by an external pump. The condensation is therefore occurring not in equilibrium, but in a nonequilibrium steady state with energy and particle flow.  The properties of such a steady state are not completely captured by an equilibrium description: in particular, whether it is condensed or not depends on the density, and therefore on the kinetics and the strength and form of pumping~\cite{marelic_experimental_2015}.

\begin{figure}[t]
    \centering
    \includegraphics[scale=1,trim=0 10 0 15,clip]{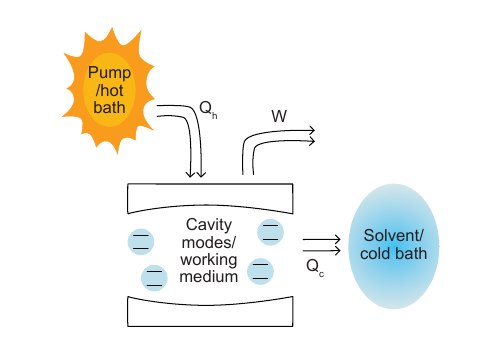}
    \caption{Photon condensation in a dye-filled cavity as a heat engine. The cavity photons which condense originate from an external source, corresponding to the hot bath of the engine, while excess energy is dissipated into the cold bath, provided by the solvent. The work output corresponds to the coherent emission from the condensate.}
    \label{fig:schematic}
\end{figure}

Here we show that there are straightforward and universal criteria for photon condensation. We focus on dye-filled cavities and investigate two different forms of pumping. The first, considered previously~\cite{klaers_boseeinstein_2010,kirton_nonequilibrium_2013,kirton_thermalization_2015}, describes the pumping of the dye with a laser at high energies, producing fluorescence which then populates the cavity modes. We focus, however, on a different situation, where the cavity modes are pumped directly, by coupling them to a thermal photon reservoir. We point out that photon condensates, similarly to lasers~\cite{scovil_three-level_1959}, fluorescent concentrators~\cite{smestad_thermodynamic_1990} and polariton condensates~\cite{toledo_tude_quantum_2024}, are forms of heat engine, which convert energy from a hot reservoir, corresponding to the pump, into work, corresponding to the coherent emission (see Fig. 1). We analyze the kinetic equations and show that the pumping strength required for condensation, encoded in the temperature of the hot reservoir $T_h$, is given by the requirement of positive entropy production. In certain limits the threshold is at zero entropy production, and corresponds to the condition for operation of a reversible (Carnot) three-level heat engine~\cite{scovil_three-level_1959, geusic_quantum_1967,toledo_tude_quantum_2024,kosloff_quantum_2014}, Eq.~\eqref{eq:revthresh}. This occurs for fast thermalization, compared with the particle gain/loss, and reservoirs which couple at a single frequency. We extend our analysis to consider pumping of the cavity modes at multiple frequencies, which leads to an increase in threshold due to entropy production and irreversibility~\cite{landi_irreversible_2021}. The increases are relatively small, however, demonstrating the feasibility of constructing a Carnot-like optical heat engine. Our results expose the thermodynamic principles of steady-state condensation and establish the general conditions under which it can be achieved. Although photon condensates have great potential as coherent light emitters and in energy harvesting and studies of quantum thermodynamics, this has been limited by the requirements of laser pumping. Our results show how this long-standing challenge could be overcome, and condensation achieved using broadband thermal sources.

For definiteness we consider a dye-filled microcavity with harmonic in-plane confinement~\cite{klaers_boseeinstein_2010}, as shown in Fig.~\ref{fig:schematic}. In such a structure there are transverse modes $m=0,1,\ldots,m_{\text{max}}$ with photon occupations $n_m$, frequencies $\omega_m=\omega_0+m\epsilon$, and degeneracies $g_m=m+1$, where $\omega_0$ is the transverse ground-state frequency, and $\epsilon$ the trapping frequency. Our analysis is based on a generalization of the kinetic equations used in~\cite{kirton_thermalization_2015,kirton_nonequilibrium_2013}:\begin{align}   
	\dot{n}_m(t)=& N_d\left[\Gamma^e_m (n_m+1) p_e  -\Gamma^a_m n_m p_g \right] \nonumber\\ & +\kappa_m \left[n^h_m(n_m +1)- (n^{h}_m+1)n_m\right].
\label{eq:nmdotmultimode}\end{align} 

The first line of Eq. \eqref{eq:nmdotmultimode} describes the emission and absorption of photons from $N_d$ dye molecules. Each is modeled as a two-state electronic system, with excited and ground state populations $p_e$ and $p_g=1-p_e$, respectively, coupled to molecular vibrations. $\Gamma_m^e$ and $\Gamma_m^a$ are the emission and absorption rates. These can be computed from microscopic models. However, for our purposes it suffices that they satisfy the Kennard-Stepanov relation~\cite{kennard_thermodynamics_1918} $\Gamma_m^e=\Gamma_m^a e^{-(\omega_m-\omega_d)/T_c}$, where $\omega_d$ is the zero-phonon transition frequency, and $T_c$ the temperature of the solvent ($\hbar=k=1$). The electronic population obeys~\begin{equation}
    \dot{p}_e(t)=-({\Gamma}_{\downarrow}^{\rm tot}+\pumpdn) p_e+ ({\Gamma}_{\uparrow}^{\rm tot}+\pumpup)p_g.\label{eq:pedot}\
\end{equation} Here $\Gamma_\downarrow^{\rm tot}=\sum_m g_m \Gamma^e_m (n_m+1)$ and $\Gamma_\uparrow^{\rm tot}=\sum_m g_m \Gamma^a_m n_m$ are transition rates due to the emission and absorption of cavity photons, respectively, while $\pumpdn$ and $\pumpup$ describe other processes.

We consider two different ways in which system can be pumped. In the first scenario, considered previously~\cite{kirton_thermalization_2015, kirton_nonequilibrium_2013}, the dye is pumped and fluoresces into the cavity modes, creating photons which are lost through the mirrors with rates $\kappa_m$. In this case we set $\pumpup, \pumpdn \neq 0$ in Eq. (\ref{eq:pedot}) to account for excitation and de-excitation by the pump. The loss of cavity photons is described by the second line of Eq.~\eqref{eq:nmdotmultimode} with $n_m^h=0$. In the second scenario, illustrated in Fig. 1, the cavity modes are pumped, by coupling them to an external thermally-populated photon reservoir with temperature $T_h$. The second line of Eq.~\eqref{eq:nmdotmultimode} describes the exchange of photons with such a reservoir, with each cavity mode coupled to a corresponding reservoir mode of population $n^{h}_m=1/(e^{\omega_m/(kT_h)}-1)$. This corresponds to the situation where a thermal source, such as sunlight with $T_h\approx 6000\;\mathrm{K}$, is imaged onto the cavity, providing both a source (first term) and sink (second term) of particles for condensation. As there is no direct pumping of the dye $\pumpup=0$ and $\pumpdn$ is negligible. Note that we are considering the case where the source is coupled to the long-lived cavity modes in regions of high mirror reflectivity. Thus the loss rates $\kappa_m$ are small for all relevant modes; we take them to be equal $\kappa_m=\kappa$. This approach differs from a recent sunlight-pumped experiment~\cite{busley_sunlight-pumped_2023}, where the pump was incident at a high angle, where the mirror reflectivity was low.

We begin by giving a thermodynamic interpretation of the threshold in the dye-pumped case. For each reservoir in Eq.\ \eqref{eq:nmdotmultimode}, for example the dye, we have a gain term, proportional to $(n_m+1)$, and a loss term, proportional to $n_m$. The ratio of the prefactors, $\gamma^e_m$ and $\gamma^a_m$,  defines an effective temperature for each mode,\begin{equation}\frac{\gamma^e_m}{\gamma^a_m}=e^{-\omega_m/T^{\rm eff}_m}.\label{eq:efftemp}\end{equation} This is the temperature of a thermal bosonic reservoir that is equivalent to the dye. If only the dye were present the steady-state population is $n_m=(\gamma^e_m/\gamma^a_m-1)^{-1}$, and the mode is in equilibrium at $T^{\rm eff}_m$ (and $\mu=0$ as in Eq.~\eqref{eq:efftemp}). More generally, however, the steady-state population and effective temperature will be intermediate between those produced by the different reservoirs.

For the coupling to the dye we have \begin{equation}\frac{\gamma_m^e}{\gamma_m^a}=\frac{\Gamma_m^e p_e}{\Gamma_m^a p_g}=\frac{\Gamma_m^e (\Gamma_\uparrow^{\mathrm{tot}}+\pumpup)}{\Gamma_m^a (\Gamma_\downarrow^{\mathrm{tot}}+\pumpdn)}.\end{equation} The pump excites transitions of energy $\omega_d$, so its equivalent temperature $T^{\text{dye}}_h$ obeys $\pumpup/\pumpdn=e^{-\omega_d/T^{\text{dye}}_h}$. Using this relation, conservation of total number of particles in the steady state $\sum_m g_m \dot{n}_m = 0 $, and the Kennard-Stepanov relation, we find
\begin{equation}
    \frac{\gamma^e_m}{\gamma^a_m}= e^{-(\omega_m-\omega_d)/T_c}\left(e^{-\omega_d/T_h}- \frac{N_{\rm ph} \ \kappa}{N_d \ p_g \ {\Gamma}_{\downarrow}}\right),\label{eq:rates}
\end{equation} where $N_{\rm ph}=\sum g_m n_m$ is the total photon number. We neglect the second term in the fast-thermalization limit $\kappa\rightarrow 0$. The effective temperature of the ground-state mode is then given by\begin{equation}
\frac{\omega_0}{T^{\mathrm{eff}}_0}=\frac{\omega_d}{T^{\text{dye}}_h}+\frac{\omega_0-\omega_d}{T_c}.
\label{eq:2ndlaw}
\end{equation} 

In the limit $\kappa\rightarrow 0$ the threshold pump temperature $T^{\text{dye}}_h$ for condensation is given by Eq.~\eqref{eq:2ndlaw} with $T^{\mathrm{eff}}_0\rightarrow \infty$. This corresponds to the point where the gain from the dye reaches the absorption, $\gamma^e_0/\gamma^a_0\rightarrow 1$, and the ground-state occupation begins to diverge. Eq.~\eqref{eq:2ndlaw} then describes the entropy balance in a reversible three-level heat engine~\cite{kosloff_quantum_2014,scovil_three-level_1959,geusic_quantum_1967,geva_quantum_1996,geva_three-level_1994,mitchison_quantum_2019,landi_irreversible_2021}, with quantized transition energies $\omega_d$ and $\omega_d-\omega_0$ associated with the hot and cold reservoirs respectively. The first term on the right is the entropy increase of the working medium when it absorbs heat $\omega_d$ from the pump to excite the dye, and hence the cavity modes, while the second is the entropy decrease as the photon is transferred to the ground state mode, and the heat $\omega_d-\omega_0$ flows to the solvent. In an ideal condensate this increases the population of the macroscopically occupied state at $\omega_0$, which produces negligible entropy increase, and so is associated with work~\cite{toledo_tude_quantum_2024}. Thus condensation is only possible if the entropy changes at the cold and hot reservoir balance. More generally, we can interpret the left-hand side of Eq. (\ref{eq:2ndlaw}) as an entropy change associated with the effective temperature of the mode. Our approach is consistent with the statement that work is energy supplied without any associated entropy increase and described by an infinite temperature reservoir~\cite{geusic_quantum_1967}. When $\kappa \neq 0$ the threshold requires that the emission into the condensate mode not only balances the absorption but also the decay of the cavity mode, so $\gamma^e_0/\gamma^a_0 > 1$, and the threshold is at $T^{\mathrm{eff}}_0<0$. This is analogous to the requirement of inversion in a laser, and can be seen, from Eq. (\ref{eq:2ndlaw}), to increase (decrease) the hot (cold) bath temperature at threshold compared with the reversible limit $\kappa\rightarrow 0$. The last term in the bracket in Eq.\ \eqref{eq:rates} represents a loss of particles into modes other than the condensate, which will further shift the threshold temperatures.

We now turn to the second scenario, where there is no direct pumping of the electronic transitions ($\pumpup = \pumpdn = 0$), and the pump is provided by a thermal photon reservoir coupled to the cavity modes ($n_m^h\neq 0$). To analyze this case we use the fact that the scattering with the dye produces a Bose-Einstein distribution with temperature $T_c$ and chemical potential $\mu=\omega_d+k T_c\ln p_e/p_g$\cite{klaers_boseeinstein_2010,klaers_statistical_2012,kirton_thermalization_2015,kirton_nonequilibrium_2013}. This holds because, from the Kennard-Stepanov relation, we have \begin{equation}\frac{\gamma_m^e}{\gamma_m^a}=\frac{\Gamma_m^e p_e}{\Gamma_m^a p_g}= e^{- (\omega_m-\mu)/T_c}.\label{eq:ratesanddistn}\end{equation} Note that here the ratio of the rates is being used to define an effective temperature and chemical potential for the dye, whereas in Eq.~\eqref{eq:efftemp} it was used to define only a temperature. These different definitions give different temperatures but the same results.

The external thermal photon reservoir, in the absence of the dye, would drive the cavity modes to another equilibrium state, with temperature $T_h$ and $\mu=0$. However, because $\kappa\ll\Gamma_m^e,\Gamma_m^a$, the distribution is accurately approximated by that produced by the dye. The external reservoir nonetheless plays an important role, determining the total population and hence the chemical potential in the steady-state. Since it is the only source or sink of particles, the total flux between it and the cavity modes vanishes in the steady-state, which implies that the total photon number in the cavity is the same as that in the corresponding reservoir states, \begin{equation}\sum_m g_m n_m=\sum_m g_m n_m^h. \label{eq:fluxbalance} \end{equation} 

We can obtain a threshold corresponding to the reversible (Carnot) limit, Eq.~\eqref{eq:2ndlaw}, if the cavity modes couple to the external thermal reservoir at only two energies, so that only two terms appear in the sums in Eq.~\eqref{eq:fluxbalance}. One energy level should correspond to the ground state with energy $\omega_0$, degeneracy $g_0$, which we take to be one, and cavity photon population $n_0$. The other corresponds to the excited states with energy $\omega_s$, degeneracy $g_s \gg g_0$, and cavity photon population per mode $n_s$. Since $g_s \gg g_0$ we can neglect the terms with $m=0$ in Eq.\ \eqref{eq:fluxbalance}. Using the respective Bose functions for the external and internal populations in the excited states, $n_s=1/(e^{(\omega_s-\mu)/T_c}-1), n_s^h=1/(e^{\omega_s/T_h}-1)$ and setting $\mu=\omega_0$, gives the condensation threshold \begin{equation} 0 = \frac{\omega_s}{T_h}+\frac{\omega_0-\omega_s}{T_c}. \label{eq:revthresh} \end{equation} This can be seen to be the same form obtained for the threshold in the dye-pumped case, Eq.\ \eqref{eq:2ndlaw}, modified because the energy absorbed from the hot bath, in the first terms on the right-hand sides, is now $\omega_s$ rather than $\omega_d$, changing also the energy emitted to the cold bath, appearing in the second terms. 

\begin{figure}[b]
    \centering
    \includegraphics[scale=0.5]{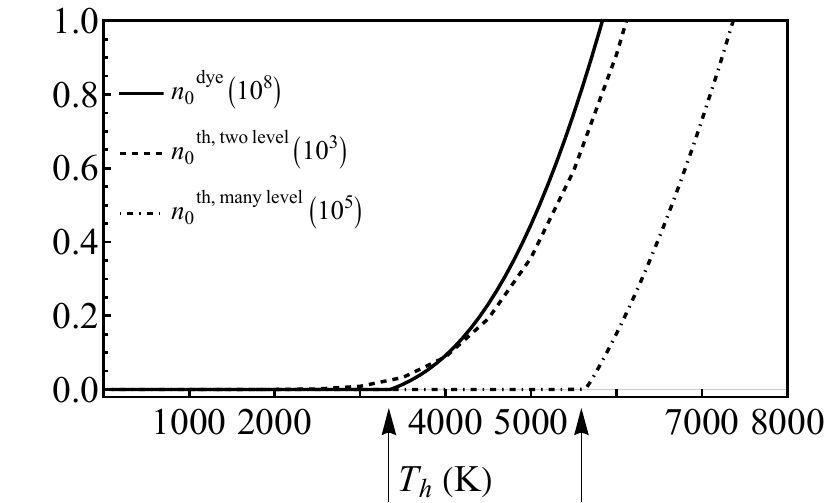}
    \caption{Ground-state photon occupations as functions of the pump strength expressed as a temperature $T_h$, for different forms of pumping. Solid : pumping of the dye, with $T_h$ the temperature of its electronic transition. Dotted and dot-dashed: pumping of the cavity modes by coupling to an external thermal photon reservoir of temperature $T_h$, considering only two cavity energy levels (dotted) or all of them (dot-dashed). The thresholds for dye pumping and the two-level cavity agree with Eq. \eqref{eq:revthresh} (left arrow). The threshold for many-level thermal pumping agrees with Eq. \eqref{eq:threshold} (right arrow).}
    \label{fig:particlenumber}
\end{figure}

In Fig. \ref{fig:particlenumber} we show how these predictions compare with the results of a simulation of Eqs.\ \eqref{eq:nmdotmultimode} with realistic dye spectra and other parameters~\cite{klaers_boseeinstein_2010,busley_sunlight-pumped_2023}. The emission and absorption rates, $\Gamma_m^e, \Gamma_m^a$, are computed using the expressions in~\cite{kirton_thermalization_2015,kirton_nonequilibrium_2013}, with parameters chosen to fit the rhodamine-6G spectrum~\cite{schmitt_absorption_2024}\footnote{ We use $S=0.5, \Omega=5~\mathrm{ps}^{-1}, \gamma=2\pi \times 50\,\mathrm{ps}^{-1}, g=7~\mathrm{ns}^{-1}$, in the notation of \cite{kirton_thermalization_2015}}. We take $\omega_d/2\pi=556$ THz, $\epsilon/2\pi = 40$ GHz, and $(\omega_0-\omega_d)/2\pi=-50$ THz. The solid black curve shows the ground-state population obtained in the dye-pumped scenario, where we include $m_{\mathrm{max}}+1=3700$ levels. The dashed black curve is the corresponding result for the simplified thermally pumped case, where as above we include only two energy levels. We take the higher one to be at $\omega_s=\omega_d$, so that the predicted condensation thresholds, $T_h=T_c\omega_s/(\omega_s-\omega_0)$, are the same in both cases. The corresponding value is marked with the black arrow, and can be seen to be in perfect agreement with the numerical results. It may be noted that, although the condensation thresholds in the dye-pumped and thermally-pumped cases are the same, for this choice of $\omega_s$,
 the occupation numbers differ greatly. This difference reflects the number of states pumped by the hot reservoir, which is $N_d=10^9$ and $g_s=7.5\times 10^{4}$ in the two cases shown.

\begin{figure}[t]
    \centering
    \includegraphics[scale=0.5]{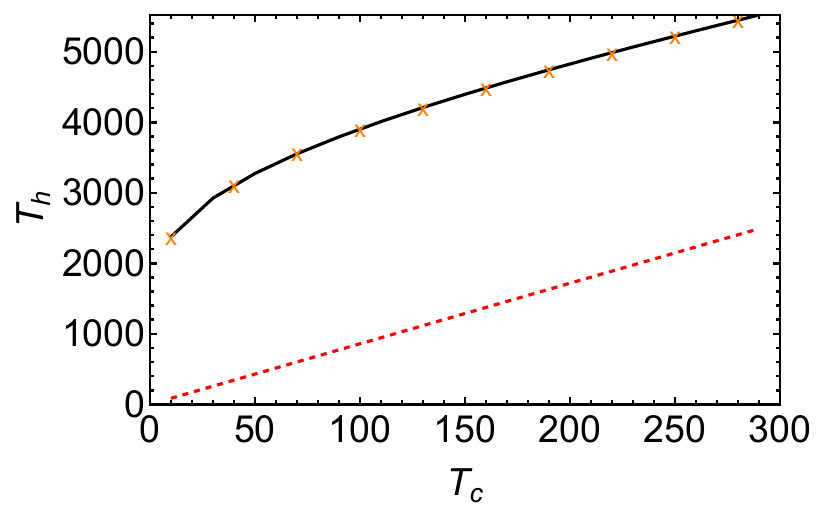}
    \caption{Phase diagram of photon condensation in a many-level harmonic trap with external thermal pumping, calculated using Eq.\ \eqref{eq:threshold} (black curve) and by numerically evolving Eq.\ \eqref{eq:nmdotmultimode} (orange points). The red dashed curve is the corresponding threshold for a reversible heat engine. Regions above the curves are condensed.}
    \label{fig:phasediagram}
\end{figure}

We now extend our analysis to consider a thermal reservoir coupled to all the levels of a harmonic trap. Simulation results showing the occupation of the ground-state mode are given in Fig.~\ref{fig:particlenumber} for this case, showing the threshold $T_h$ where condensation appears. To calculate this threshold we take the continuum limit, $\epsilon\rightarrow 0$ of Eq.~\eqref{eq:fluxbalance} and set $\mu=\omega_0$, to find \begin{equation} \int_{\omega_0}^{\omega_{\mathrm{max}}} d\omega \frac{\rho(\omega) }{e^{\beta_c(\omega-\omega_0)}-1}=\int_{\omega_0}^{\omega_{\mathrm{max}}}  d\omega \frac{\rho(\omega)}{e^{\beta_h\omega}-1},\label{eq:threshold}\end{equation} with the density of states $\rho(\omega)=(\omega-\omega_0)/(2\epsilon^2)$. These integrals can be related to the polylogarithm function (Bose-Einstein integral), with that on the left giving the critical particle number for condensation in a harmonic trap, $N_{\mathrm{th}}=\pi^2 T_c^2/(6\epsilon^2)$ when $\omega_{\mathrm{max}}\rightarrow \infty$. The critical temperature obtained by solving Eq.~\eqref{eq:threshold} for $T_c=300$ K is marked in Fig.~\ref{fig:particlenumber}. It agrees with the numerics, and is of the same order-of-magnitude as that for the reversible two-level/dye-pumped case with input energy $\omega_s=\omega_d$. Fig.~\ref{fig:phasediagram} shows how the threshold $T_h$ varies with $T_c$. For comparison, we show, with the red line, the threshold for the reversible heat engine, Eq.~\eqref{eq:revthresh}, with the input energy $\omega_s$ taken as the mean energy per particle of the hot reservoir, $\bar\omega_h$. 

\begin{figure}
    \centering
    \includegraphics[width=.49\textwidth]{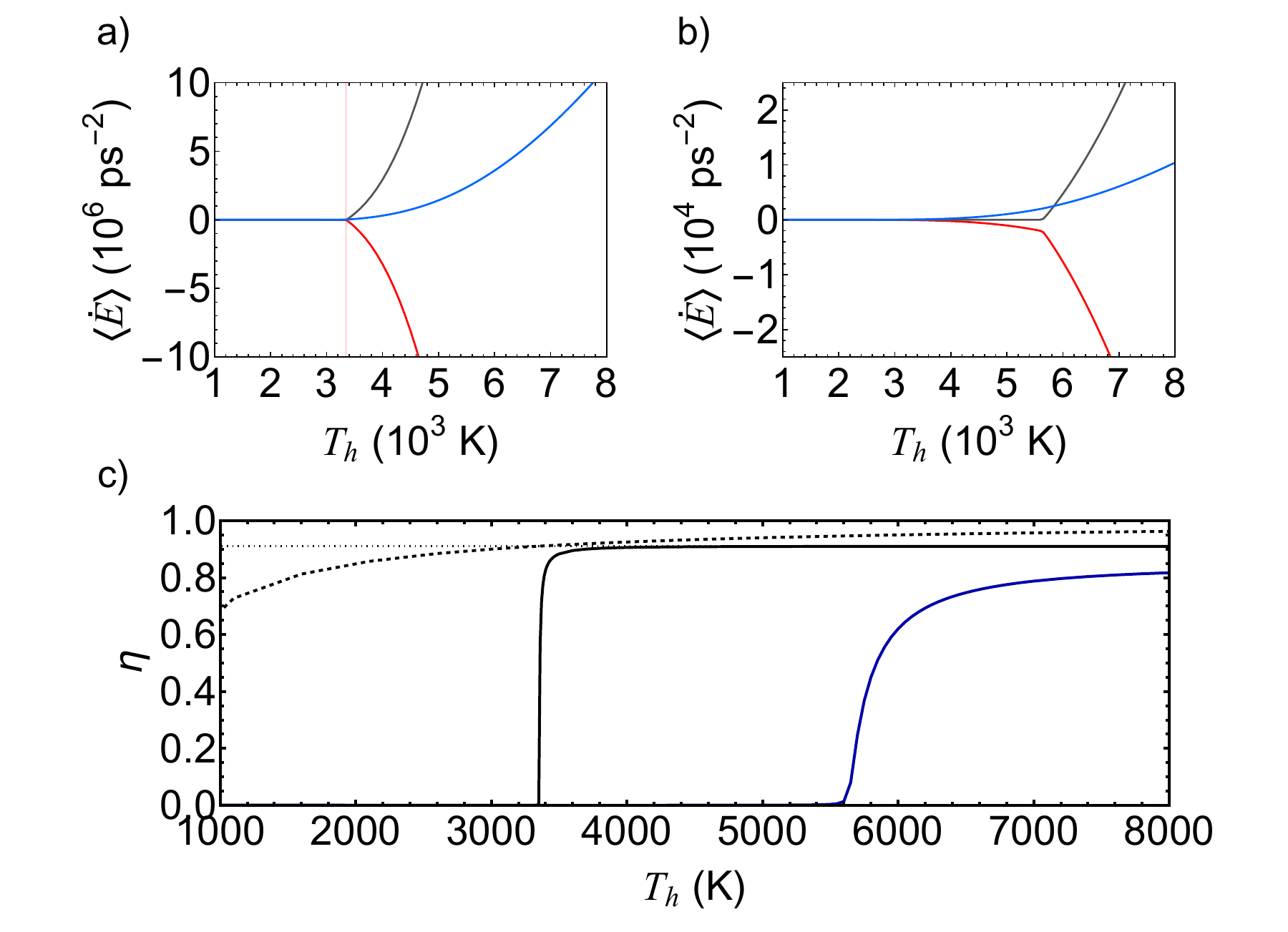}
    \caption{(a,b) Energy currents for dye pumping (a) and external thermal pumping of a many-level trap (b). The black curves are the work outputs, and the red (blue) curves the currents to the hot (cold) bath. (c) Efficiencies for dye pumping (black solid) and external pumping (blue solid). For dye pumping the efficiency is determined by the energy ratio $\omega_0/\omega_d$ (horizontal dashed line). The threshold is where this crosses the Carnot efficiency for the given bath temperatures (black dashed curve).}
    \label{fig:heatandefficiency}
\end{figure}

In Fig.~\ref{fig:heatandefficiency} we show that heat and work currents and heat-engine efficiencies for the dye-pumped and thermally-pumped cases. In the dye-pumped case there are no currents below threshold, so the working medium decouples into one part in equilibrium with the hot bath and one in equilibrium with the cold. The heat and work currents appear together at threshold, but as they are infinitesimal at that point, the heat transfers to the baths are reversible, so that the entropy changes correspond to Eq.~\eqref{eq:2ndlaw}. The threshold occurs with increasing $T_h$ when the left-hand-side of Eq.~\eqref{eq:2ndlaw} is zero. As can be seen in Fig.~\ref{fig:heatandefficiency}(c), this is the point where the Carnot efficiency for the given bath temperatures (dashed curve) crosses the fixed efficiency of the device given by the ratio of the quantized output to input energy $\omega_0/\omega_d$ (horizontal line). The same physics gives the threshold of the three-level laser~\cite{scovil_three-level_1959}, and our two-level thermally pumped model. The efficiency actually achieved, in the dye-pumped case, can be seen to rise rapidly at threshold to a value slightly below these, before curving to approach the fixed efficiency as $T_h$ increases. These departures from the ideal step-like form are due to the non-zero loss rate $\kappa$, in comparison to the absorption, $N_d\Gamma_0^a$.

Fig. \ref{fig:heatandefficiency}(b) shows the currents for the many-level thermal pump. In comparison to Fig.\ \ref{fig:heatandefficiency}(a) there are now energy flows from the hot to the cold bath below threshold. The cavity modes (working medium) interact with both reservoirs at threshold, so will not be in equilibrium with either separately. Thus there will be entropy generation in the heat exchanges, which will tend to increase (decrease) the critical $T_h$ ($T_c$) for condensation. Furthermore, due to these heat currents the efficiency is now zero at threshold, and increases smoothly with $T_h$. For large $T_h$ it approaches a limiting value corresponding to $\omega_0/\bar\omega_{h}$, since the emission from the condensate then dominates over the other dissipation channels. 

In conclusion, we have shown how photon condensation can be understood from the perspective of non-equilibrium thermodynamics. Condensation can occur not only when a laser is used to excite the dye at high energies, creating fluorescence which populates the cavity modes, but also when those cavity modes are populated directly through their coupling to a thermal source. We have shown that the threshold temperatures for condensation are determined by the entropy balance in a heat-engine cycle, and correspond to those of a reversible (Carnot) three-level engine in the case of spectrally narrow baths. The threshold is larger for a multimode thermal source, but of the same order of magnitude. Our results show that condensation could be achieved using broadband sources, allowing the realization of optical heat engines which convert broadband to narrowband light. This may find applications in coherent light emission and energy harvesting~\cite{rau_efficiency_2005,meyer_chemical_2009,busley_sunlight-pumped_2023,van_sark_luminescent_2008,smestad_thermodynamic_1990}. It may also prove useful as an experimental setting for studies of quantum thermodynamics, probing, for example, the impact of quantum coherence and correlations~\cite{vinjanampathy_quantum_2016} and thermodynamic uncertainty relations~\cite{horowitz_thermodynamic_2020,kalaee_violating_2021}. 

Code and data supporting this article are openly available~\cite{data}. 

\begin{acknowledgments}
We acknowledge funding from Taighde \'Eireann -- Research Ireland (21-FF-P/10142). 
\end{acknowledgments}


%

\end{document}